\documentclass[twocolumn]{aastex63}
\usepackage{graphicx}
\usepackage{enumerate}
\usepackage{amssymb, amsmath}
\usepackage{natbib}
\usepackage{color}
\usepackage{ulem}
\usepackage{dblfnote}
\usepackage{appendix}
\usepackage{hyperref}
\usepackage[stable]{footmisc}
\usepackage{relsize}

\definecolor{red}{rgb}{1.0,0.0,0.0}

\begin{document}

\title{SCExAO/CHARIS Direct Imaging of A Low-Mass Companion At A Saturn-Like Separation from an Accelerating Young A7 Star}

\email{jchilcote@nd.edu}
\author{Jeffrey Chilcote}
\affiliation{Department of Physics, University of Notre Dame, Notre Dame, IN, USA}
\author{Taylor Tobin}
\affiliation{Department of Physics, University of Notre Dame, Notre Dame, IN, USA}
\author{Thayne Currie}
\affiliation{Subaru Telescope, National Astronomical Observatory of Japan, 
650 North A`oh$\bar{o}$k$\bar{u}$ Place, Hilo, HI  96720, USA}
\affiliation{NASA-Ames Research Center, Moffett Blvd., Moffett Field, CA, USA}
\affiliation{Eureka Scientific, 2452 Delmer Street Suite 100, Oakland, CA, USA}
\author{Timothy D. Brandt}
\affiliation{Department of Physics, University of California, Santa Barbara, Santa Barbara, California, USA}
\author{Tyler D. Groff}
\affiliation{NASA-Goddard Space Flight Center, Greenbelt, MD, USA}
\author{Masayuki Kuzuhara}
\affil{Astrobiology Center of NINS, 2-21-1, Osawa, Mitaka, Tokyo, 181-8588, Japan}
\author{Olivier Guyon}
\affiliation{Subaru Telescope, National Astronomical Observatory of Japan, 
650 North A`oh$\bar{o}$k$\bar{u}$ Place, Hilo, HI  96720, USA}
\affil{Steward Observatory, The University of Arizona, Tucson, AZ 85721, USA}
\affil{College of Optical Sciences, University of Arizona, Tucson, AZ 85721, USA}
\affil{Astrobiology Center of NINS, 2-21-1, Osawa, Mitaka, Tokyo, 181-8588, Japan}
\author{Julien Lozi}
\affiliation{Subaru Telescope, National Astronomical Observatory of Japan, 
650 North A`oh$\bar{o}$k$\bar{u}$ Place, Hilo, HI  96720, USA}
\author{Nemanja Jovanovic}
\affiliation{Department of Astronomy, California Institute of Technology, 1200 East California Boulevard, Pasadena, CA 91125}
\author{Ananya Sahoo}
\affiliation{Subaru Telescope, National Astronomical Observatory of Japan, 
650 North A`oh$\bar{o}$k$\bar{u}$ Place, Hilo, HI  96720, USA}
\author{Vincent Deo}
\affiliation{Subaru Telescope, National Astronomical Observatory of Japan, 
650 North A`oh$\bar{o}$k$\bar{u}$ Place, Hilo, HI  96720, USA}

\author{Eiji Akiyama}
\affiliation{Department of Engineering, Niigata Institute of Technology, 1719 Fujihashi, Kashiwazaki, 945-1195, Japan}
\author{Markus Janson}
\affiliation{Department of Astronomy, Stockholm University, AlbaNova University Center, 106 91 Stockholm, Sweden}
\author{Jill Knapp}
\affiliation{Department of Astrophysical Science, Princeton University, Peyton Hall, Ivy Lane, Princeton, NJ 08544, USA}
\author{Jungmi Kwon}
\affiliation{Department of Astronomy, Graduate School of Science, The University of Tokyo, 7-3-1, Hongo, Bunkyo-ku, Tokyo, 113-0033, Japan}
\author{Michael W. McElwain}
\affiliation{NASA-Goddard Space Flight Center, Greenbelt, MD, USA}
\author{Jun Nishikawa}
\affiliation{Department of Astronomy, Graduate School of Science, The University of Tokyo, 7-3-1, Hongo, Bunkyo-ku, Tokyo, 113-0033, Japan}
\affil{Department of Astronomical Science, School of Physical Sciences, The Graduate University for Advanced Studies (SOKENDAI), 2-21-1, Osawa, Mitaka, Tokyo, 181-8588, Japan}
\affil{Astrobiology Center of NINS, 2-21-1, Osawa, Mitaka, Tokyo, 181-8588, Japan}
\author{Kevin Wagner}
\affil{Steward Observatory, The University of Arizona, Tucson, AZ 85721, USA}
\altaffiliation{NASA Hubble Fellow}
\author{Krzysztof Hełminiak}
\affil{Nicolaus Copernicus Astronomical Center, Polish Academy of Sciences, ul. Rabianska 8, 87-100 Torun, Poland}
\author{Motohide Tamura}
\affil{Astrobiology Center of NINS, 2-21-1, Osawa, Mitaka, Tokyo, 181-8588, Japan}
\affiliation{Department of Astronomy, Graduate School of Science, The University of Tokyo, 7-3-1, Hongo, Bunkyo-ku, Tokyo, 113-0033, Japan}
\affiliation{National Astronomical Observatory of Japan, 2-21-2, Osawa, Mitaka, Tokyo 181-8588, Japan}

\shortauthors{Chilcote et al.}
\begin{abstract}
We present the SCExAO direct imaging discovery and characterization of a low-mass companion to the nearby young A7IV star, HD 91312.  SCExAO/CHARIS $JHK$ (1.1--2.4 $\mu m$) spectra and SCExAO/HiCIAO $H$ band imaging identify the companion over a two year baseline in a highly inclined orbit with a maximum projected separation of 8 au. The companion, HD 91312~B, induces an 8.8-$\sigma$ astrometric acceleration on the star as seen with the Gaia \& Hipparcos satellites and a long-term radial velocity trend as previously identified by \citet{Borgniet2019}. HD 91312 B's spectrum is consistent with that of an early-to-mid M dwarf. Hipparcos and Gaia absolute astrometry, radial-velocity data, and SCExAO/CHARIS astrometry constrain its dynamical mass to be $0.337^{+0.042}_{-0.044}$~M\textsubscript{\(\odot\)}, consistent with - but far more precise than - masses derived from spectroscopy, and favors a nearly edge-on orbit with a semi-major axis of $\sim$9.7 au. This work is an example of precisely characterizing properties of low-mass companions at solar system-like scales from a combination of direct imaging, astrometry, and radial-velocity methods.

\end{abstract}

\section{Introduction}

Direct imaging and two \textit{indirect} methods -- radial-velocity (RV) and astrometry -- provide ways to detect and characterize young exoplanets, brown dwarfs, and low-mass stellar companions at Jupiter or greater separations around nearby stars \citep{Marois2008a,Nakajima1995,Zimmerman2010,Borgniet2019,Lagrange2020,Currie2020b,Brandt2018,Brandt2019}. Each method in isolation has its strengths and weaknesses.  While direct imaging can probe key atmospheric properties -- e.g. temperature, clouds, chemistry, and gravity -- and provide some constraints on orbits, the masses this method derives are inferred from luminosity evolution models which themselves are uncertain. Furthermore, astrometric coverage from high-contrast direct imaging is usually small compared to the companion's likely orbital period \citep{Barman2015,Currie2011,Spiegel2012,Berardo2017,Lagrange2019,Blunt2017,Kuzuhara2013,Carson2013}. On the other hand, astrometry and RV by themselves provide constraints on orbital properties and lower limits on masses, but do not provide the same insight into atmospheric properties.

Combining direct imaging with astrometry and/or RV substantially improves our ability to characterize low-mass companions \citep{Brandt2019}. Relative astrometry of companions from imaging and absolute astrometry and/or Doppler light curves of the star can directly constrain companion masses and orbital properties. Atmospheric properties derived from imaging can be tied to the object's mass. The \textit{Hipparcos-Gaia Catalog of Accelerations} (HGCA) -- a combination of the exquisite astrometry from the \textit{Gaia} mission and those from Hipparcos -- provides a list of nearby stars whose proper motion accelerations hint at the presence of massive, imageable companions on solar system scales \citep{GAIA2018,Brandt2018}.  Direct imaging surveys targeting these accelerating stars may have significantly higher yields than blind surveys and allow substantially improved characterization capabilities \citep[e.g.][]{Calissendorff2018,Fontanive2019,Currie2020b,Bowler2021,Steiger2021}.

In this paper, we report the direct imaging detection of a low-mass companion at a projected separation of 8~au from the nearby A7IV star, HD 91312~A, using both the Coronagraphic High Angular Resolution Imaging Spectrograph \citep[CHARIS,][]{Groff2016} behind the Subaru Coronagraphic Extreme Adaptive Optics Project \citep[SCExAO,][]{Jovanovic2015} and the NIRC2 camera on Keck.  CHARIS spectra and NIRC2 photometry reveal HD 91312~B to be an early-to-mid M dwarf.  The HGCA identifies an astrometric acceleration for the primary induced by HD 91312~B; the companion also is responsible for a long-term RV drift identified by \citet{Borgniet2019}. 

\begin{deluxetable*}{llllllllll}
     \tablewidth{0pt}
    \tablecaption{HD 91312 Observing Log\label{obslog_Cap}}
    \tablehead{\colhead{UT Date} & \colhead{Instrument} &  \colhead{coronagraph} & \colhead{Seeing (\arcsec{})} &{Passband$^{a}$} & \colhead{$\lambda$ ($\mu m$)$^{a}$} 
    & \colhead{$t_{\rm exp}$ (s)} & \colhead{$N_{\rm exp}$} & \colhead{$\Delta$PA ($^{o}$)} & \colhead{PSF Subtraction} \\
    {} & {} & {} & {} & {} & {} & {} & {} & {} & \colhead{Strategy}  }
    \startdata
    20161215 & SCExAO/HiCIAO& vortex & 0.6 & $H$ & 1.65 & 30 & 79 & 44.4 &ADI\\
    20170312 & SCExAO/CHARIS & Lyot & 0.5 & $JHK$ & 1.16--2.37& 10.32 & 117 & 18.5 &ADI\\
    20170313 & SCExAO/CHARIS & Lyot & 0.6 & $JHK$ & 1.16--2.37& 14.75 & 61 & 25.5 &ADI\\
    20180208 & SCExAO/CHARIS & Lyot & 1.4 & $JHK$ & 1.16--2.37& 16.23 & 12 & 1.9 &SDI\\
    20181128 & Keck/NIRC2 & Lyot & 1.0 & $L_{\rm p}$ & 3.78 & 30 & 54 & 46.3 & ADI\\
    20181215 & SCExAO/CHARIS & none & -- & $JHK$ & 1.16--2.37& 20.65 & 35 & 34.4 &ADI\\
    \enddata
    \tablecomments{
    a) For CHARIS data, this column refers to the wavelength range.  For HiCIAO broadband imaging data, it refers to the central wavelength.
    }
    \label{obslog}
    \end{deluxetable*}

In section \ref{sec:data}, we describe the observations, data reduction, and spectral extraction obtained with SCExAO/HiCIAO, SCExAO/CHARIS, and Keck/NIRC2. In section \ref{sec:analysis}, we discuss the analysis HD~91312~B. We find that both an empirical spectral comparison and a comparison to atmospheric models find a best match to an early-to-mid M dwarf. An analysis of the Gaia astrometry, orbital motion, and radial velocity trend provides a strong constraint on the mass. Finally, in section \ref{sec:stellar_evolution_models}, we compare the system to evolutionary models, providing a prediction of the age of the HD~91312 system.

\section{Target Properties, Observations, and Data}
\label{sec:data}

HD~91312~A is a bright ($\text{M}_\text{V}$=4.7) A7IV star located at a distance of $33.28\pm0.25$\,pc \citep{GAIA2018}.   \citet{Rhee2007} identify the star as an IRAS-excess source ($L_{\rm IR}/L_{\star}$ $\sim$ 10$^{-4}$) consistent with circumstellar dust, although later work has suggested this excess emission results from contamination by an unrelated background object \citep{Bulger2013}. Age estimates for the system vary, from 200 $Myr$ in \citealt{Rhee2007} to 700-900 $Myr$ or 19 $Myr$ in \citet{David2015} using Bayesian inference and isochrone fitting, respectively.  

Previous direct imaging searches failed to identify companions at separations of $\sim$ 1\arcsec{} or greater \citep[e.g.][]{Janson2013,DeRosa2014}.   However, precision radial-velocity (RV) data identify a long-term trend consistent with an unseen stellar to substellar companion orbiting beyond 5 au from the star \citep{Borgniet2019}.   Using the Hipparcos-Gaia Catalogue of Accelerations from \citet{Brandt2018} considering Hipparcos and Gaia-DR2 measurements, we identified a 2.4-$\sigma$ significant astrometric acceleration also consistent with an unseen (sub-)stellar companion.   Using updated Gaia-eDR3 astrometry, the significance increases to 8.8$\sigma$ (Brandt et al. 2021, submitted).

High-contrast imaging data for HD 91312 were obtained from SCExAO/CHARIS, Keck/NIRC2, SCExAO/HiCIAO between Dec 2016 and Dec 2018 (Table \ref{obslog}).  Our Subaru Telescope observations used the HiCIAO infrared camera \citep{Hodapp2008} in $H$ band ($\lambda_{\rm o}$ = 1.65 $\mu m$) or the CHARIS integral field spectrograph \citep{Groff2016} in broadband mode covering $JHK$ passbands simultaneously (1.16-2.37 $\mu$m, $\mathcal{R}$ $\sim$ 18) with SCExAO providing an extreme AO correction. For followup observations, we acquired Keck II observations with the NIRC2 camera in the $L_{\rm p}$ broadband filter ($\lambda_{o}$ = 3.78 $\mu m$) using Keck's facility AO system.

While conditions were photometric each night, the seeing varied substantially, ranging from $\theta_{V}$ = 0\farcs{}5--0\farcs{}6 for the March 2017 SCExAO/CHARIS data to 1--1.4\arcsec{} for the February and November 2018 SCExAO/CHARIS and Keck/NIRC2 data. Consequentially, the AO performance also varied.  Precipitable water vapor levels and the strength of telluric features prominent in channels bracketing the $JHK$ passbands for CHARIS also varied.

All CHARIS data utilized satellite spots for precise astrometric and spectrophotometric calibration \citep[e.g.][]{Jovanovic2015-astrogrids,Currie2018},  either during (2018 data) or preceding (March 2017) our main sequence of science exposures.  Except for the December 2018 epoch, we used a Lyot cornograph with a 0\farcs{}23 diameter occulting spot for all CHARIS observations. The NIRC2 data were taken without a coronagraph. All observations were conducted in ``vertical angle''/pupil tracking mode enabling angular differential imaging \citep[ADI;][]{Marois2006}. The CHARIS data also enable spectral differential imaging \citep[SDI;][]{Marois2000}.

       \begin{figure*}[ht]
    \centering
       
    \includegraphics[width=0.95\textwidth,trim=0mm 0mm 0mm 0mm,clip]{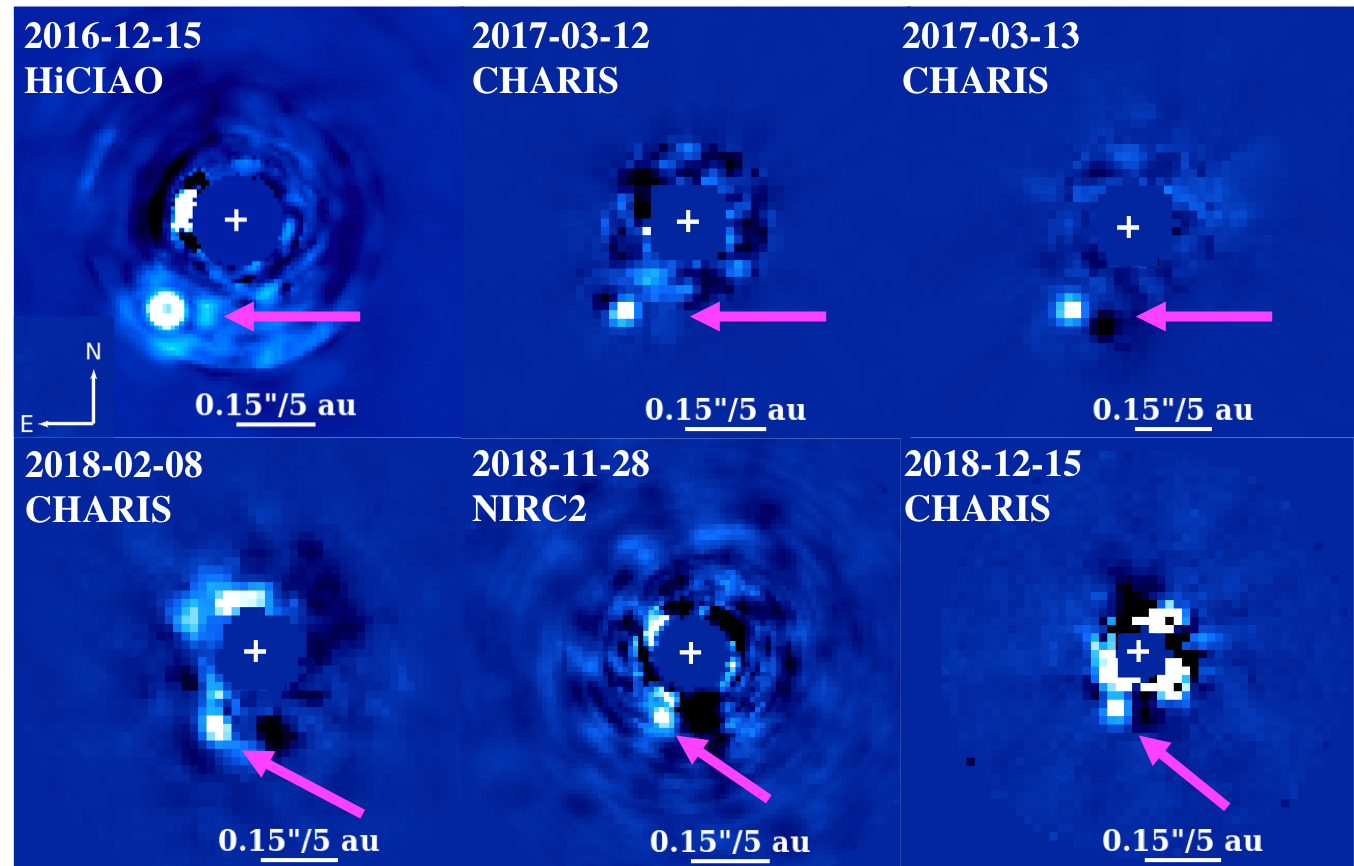}
    
    \caption{Detections of HD 91312~B (position denoted by the magenta arrow) from 2016 to 2018.}
    
    \vspace{-0.in}
    \label{fig:images}
\end{figure*}

The standard CHARIS pipeline \citep{Brandt2017} extracted data cubes from raw data; the CHARIS post-processing pipeline developed in \citet{Currie2018} provided  basic reduction steps including 
 
sky subtraction, image registration, and spectrophotometric calibration.   For spectrophotometric calibration, we adopted a Kurucz stellar atmosphere model appropriate for an A7 star. For both the HiCIAO and NIRC2 data, we used a well-tested general purpose high-contrast ADI broadband imaging pipeline \citep{Currie2010gj758,Currie2011}.

To suppress the stellar halo, we use the least-squares based algorithms -- \textit{Adaptive Locally Optimized Combination of Images} \citep[A-LOCI][]{Currie2012,Currie2015} and the \textit{Karhunen-Lo\`{e}ve Image Projection} \citep[KLIP;][]{Soummer2012-KLIP} approaches --  in combination with ADI or SDI.   We employed more conservative algorithm settings for the highest-quality data sets which also coincide with HD 91312~B's widest angular separation, adopting a high \textit{singular value decomposition} (SVD) cutoff for A-LOCI or truncating the basis set for KLIP at a small number of modes.  Due to the poorer seeing, the 2018 epoch data sets suffered from poorer AO performance and HD 91312~B at a smaller angular separation.   For these data, only aggressive approaches -- e.g. using A-LOCI in SDI mode or with a small optimization area used to construct a weighted PSF \citep[see ][]{Lafreniere2007} -- yielded a statistically significant detection. For all the reductions we used an overlap of 0.7 FWHM's. For the March 12 KLIP reduction a KL=1 was used and for the ALOCI results, we used a SVD limit of $10^-4$ was used. The March 13 ALOCI results used a SVD limit of $10^-6$. 

Figure \ref{fig:images} shows the detection of a faint point source companion, HD 91312~B, within $\rho$ $\sim$ 0\farcs{}2 of the primary.   Even with conservative algorithm settings, the signal-to-noise (SNR) for HD 91312~B's detection obtained with A-LOCI (KLIP) exceeds 68 (37) in the 2017 March 12 data.   HD 91312 B's detection significance in the 2017 March 13 data set (37) is a lower limit since a prominent negative self-subtraction footprint counterclockwise from the companion biases the true speckle noise estimate.   In the HiCIAO data, HD 91312~B slightly saturates.  No data set identifies any additional companion.   HD 91312~B appears at smaller angular separations in the 2018 data than in 2016-2017.

\section{Analysis}
\label{sec:analysis}

Prior to extracting a spectrum for HD 91312~B and determining the companion's astrometry, we corrected for signal losses due to processing by forward-modeled point sources at HD 91312~B's location using stored coefficients (for A-LOCI) or eigenvectors/eigenvalues (for KLIP) in \citet{Currie2018,Currie2020b} and \citet{Pueyo2016}. For CHARIS spectra, we focus on the 13 March 2017 data, as these have a higher throughput and weaker off-diagonal terms in their spectral covariance than the 12 March 2017 data, indicating less spatially/spectrally correlated noise \citep{GrecoBrandt2016}. The 2018 CHARIS epochs have less statistically significant detections, have likewise stronger off-diagonal terms in their spectral covariance, and/or utilized SDI-only for PSF subtraction, which complicates forward-modeling \citep{Pueyo2016}.

Our implementation of A-LOCI and KLIP only mildly attenuates HD 91312~B in the 13 March 2017 CHARIS data, as the forward-modeled companion PSF had 85-95\% throughput per channel. In the Keck/NIRC2 data, the aggressive processing needed to achieve a detection yielded far greater signal loss (23\% throughput). Except for the Keck/NIRC2 data, we found astrometric biasing to be negligible.

\subsection{Common Proper Motion, Orbit, and Dynamical Mass}
\begin{figure}
    
     \includegraphics[width=0.45\textwidth,trim=10mm 5mm 0mm 0mm,clip]{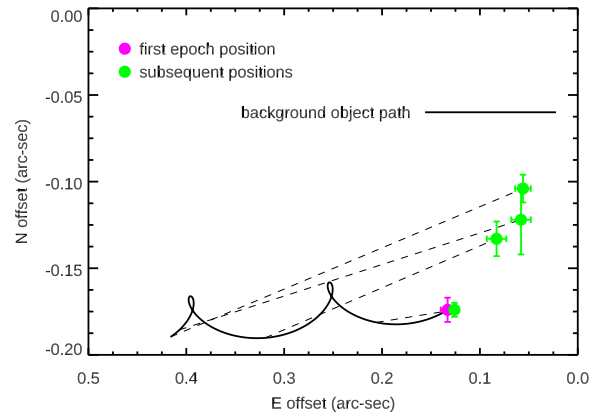}
     \vspace{-0.05in}
    \caption{Proper motion analysis for HD 91312~B.   For an initial position measured in 2016 data (magenta dot), the dotted lines connect HD 91312~B's measured position (green dots) to its predicted position were it a background object}
    \label{fig:propmot}
\end{figure}

Table \ref{astrolog} lists HD~91312~B's relative astrometry for each epoch and Figure~\ref{fig:propmot} compares the companion's positions to those expected for a background star. Gaia \& hiparcus provide absolute astrometry of the system, while the direct imaging measurements made with CHARIS, HiCIAO, and NIRC2 provide relative astrometry of the system. The primary exhibits a significant proper motion of $\Delta$\ignorespaces$\alpha$,\ignorespaces$\delta$ $\sim$ -138.1, -3.2 mas yr$^{-1}$ \citep{GAIA2018}.  HD~91312~B's astrometry over two years is therefore easily distinguishable from the path of a background star with the primary uncertainty being related to the companion SNR.

Between December 2016 and March 2017, HD~91312~B's position changes little but its angular separation is substantially smaller in subsequent epochs.   This suggests that our earliest epoch detections likely imaged the companion near its maximum projected separation.   Visual inspection of Figure~\ref{fig:propmot} shows HD 91312 B at a roughly constant position angle but varying angular separations, suggesting that its orbit is highly inclined along our line-of-sight.   

\begin{deluxetable*}{llll}
     \tablewidth{0pt}
    \tablecaption{HD 91312~B Detection Significance and Astrometry\label{detectlog}}
    \tablehead{\colhead{UT Date} & \colhead{Instrument} & \colhead{SNR$^{a}$} & \colhead{[E,N](\arcsec{})}}
    \startdata
    20161215 &SCExAO/HiCIAO & 12$^{b}$ &  [0.133, -0.174] $\pm$ [0.007, 0.007]\\
    20170312 &SCExAO/CHARIS & 68 & [0.126, -0.176] $\pm$ [0.004, 0.004]\\
    20170313 &SCExAO/CHARIS & 37$^{c}$ & [0.127, -0.172] $\pm$ [0.004, 0.004]\\
    20180208 &SCExAO/CHARIS & 6.3 & [0.083, -0.133] $\pm$ [0.010, 0.010]\\
    20181128 &Keck/NIRC2 & 3.1 & [0.058, -0.122] $\pm$ [0.010, 0.020]\\
    20181215 &SCExAO/CHARIS & 5.3 & [0.056, -0.104] $\pm$ [0.008, 0.008]
    \enddata
    \tablecomments{a) All HD 91312~B SNR estimates draw from reductions used to calculate astrometry. b) This is a lower limit since HD 91312~B is saturated in HiCIAO images. c) The SNR is likely underestimated, as negative self-subtraction footprints bias the true speckle noise estimate.
    }
    \label{astrolog}
    \end{deluxetable*}
    
\subsection{The Atmosphere of HD 91312~B}
\begin{figure}
    
     \includegraphics[width=0.45\textwidth,trim=10mm 2mm 0mm 0mm,clip]{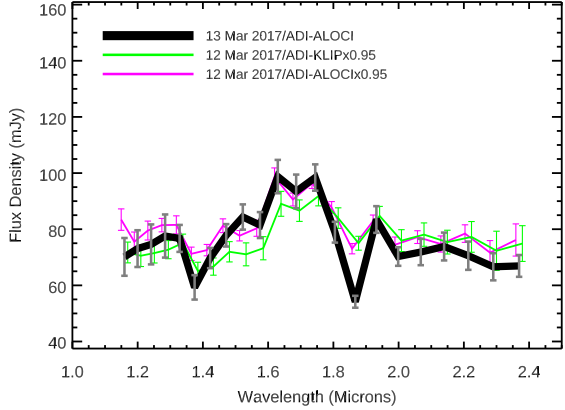}
     \vspace{-0.05in}
    \caption{CHARIS spectrum extracted from 13 March 2017 data reduced with ADI/A-LOCI (black thick line, gray error bars) compared to spectra extracted on different nights or with different processing, scaled to roughly match the 13 March 2017 spectrum's absolute flux density level. Errors are draw from the intrinsic SNR of the detection and the uncertainty in spectrophotometric calibration at that given channel.}
    \label{fig:spec}
\end{figure}

The HD 91312~B spectrum shows peaks at $J$ and especially $H$ band characteristic of M and L-type companions \citep[e.g.][]{Gagne2015} (Figure \ref{fig:spec}). The 13 March 2017 data show slightly discrepant measurements at 1.85 $\mu m$, likely due to that night's stronger telluric emission between \textit{J,H,K} bands.   HD 91312~B's 
broadband photometry in standard Maunakea Observatory filters derived from the 13 March 2017 CHARIS and 28 November 2018 NIRC2 data is $J = 10.79 \pm 0.10$, $H = 10.13 \pm 0.06$, $K_{\rm s} = 9.93 \pm 0.07$, and $L_{\rm p} = 9.79 \pm 0.35$.

\begin{figure}
    
     \includegraphics[width=0.45\textwidth,trim=0mm 0mm 0mm 0mm,clip]{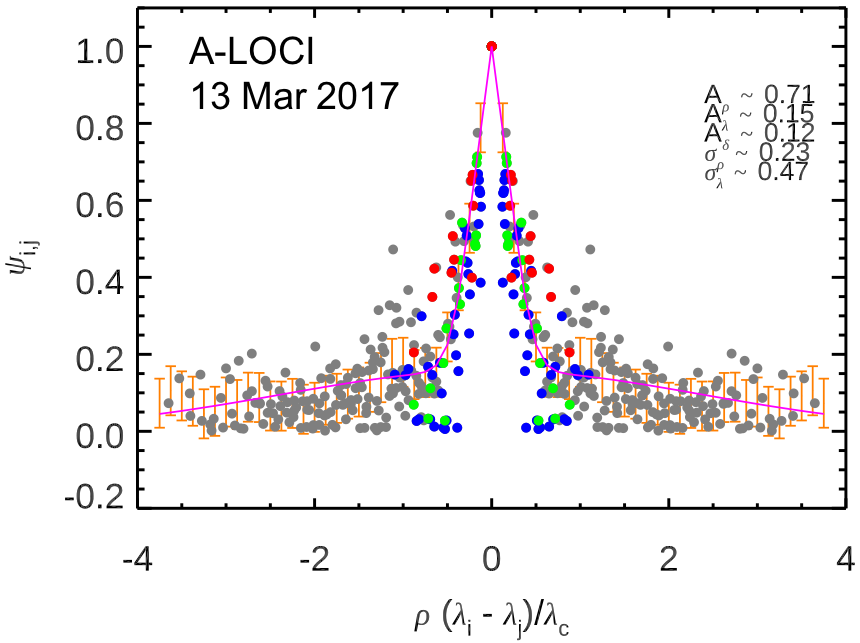}
     \vspace{-0.15in}
    \caption{Spectral covariance at the angular separation of HD 91312~B for the A-LOCI-reduced 2017 March 13 data.  The magenta line shows our fit to the spectral covariance as a function of scaled separation -- $\rho$($\lambda_{\rm i}$-$\lambda_{\rm j}$)/$\lambda_{\rm c}$ -- where $\rho$ is the separation in $\lambda$/D units for the central wavelength $\lambda_{\rm c}$ \citep[see ][]{GrecoBrandt2016}.   Blue, red, and green circles denote individual measurements between channels within the same major near-IR filter ($J$, $H$, or $K_{\rm s}$) while grey circles denote other individual measurements.   Orange points with error bars denote binned averages with 68\% confidence intervals.   The channel-independent noise contribution is even smaller for the 12 March 2017 A-LOCI and KLIP reductions.}
    \label{fig:covar}
\end{figure}

In contrast to results for wider-separation companions like HD 33632 Ab \citep{Currie2020b}, the spectral covariance at HD 91312~B's location includes substantial off-diagonal terms (Figure \ref{fig:covar}), especially for spatially-correlated noise (A$_{\rm \rho}$ $\sim$ 0.71).   More aggressive ADI processing or employing SDI would reduce the spatially/spectrally correlated noise but would complicate forward-modeling and thus impact the fidelity of our extracted spectrum.   

\begin{deluxetable}{llll}
     \tablewidth{0pt}
    \tablecaption{HD 91312 B Spectrum}
    \tablehead{\colhead{Wavelength ($\mu$m)} & \colhead{$F_{\rm \nu}$ (mJy)} &  \colhead{$\sigma$~$F_{\rm \nu}$ (mJy)} & \colhead{SNR}}
    \startdata
1.1596 & 70.1401 & 6.7327 & 10.4178 \\
1.1997 & 73.0861 & 6.5475 & 11.1625 \\
1.2412 & 74.5914 & 7.1210 & 10.4748 \\
1.2842 & 77.5707 & 7.6822 & 10.0975 \\
1.3286 & 76.7243 & 4.7999 & 15.9847 \\
1.3746$^{a}$ & 59.2881 & 4.3527 & 13.6211 \\
1.4222$^{a}$ & 69.7233 & 3.6030 & 19.3515 \\
1.4714 & 77.8220 & 3.9974 & 19.4682 \\
1.5224 & 84.3496 & 4.4957 & 18.7621 \\
1.5750 & 81.4769 & 4.6191 & 17.6391 \\
1.6296 & 98.7482 & 5.9409 & 16.6217 \\
1.6860 & 93.4650 & 5.9960 & 15.5878 \\
1.7443 & 98.4082 & 4.6971 & 20.9508 \\
1.8047 & 78.9408 & 3.7200 & 21.2204 \\
1.8672$^{a}$ & 54.1538 & 2.1265 & 25.4661 \\
1.9318$^{a}$ & 83.4531 & 4.7505 & 17.5671 \\
1.9987 & 70.2701 & 3.3121 & 21.2159 \\
2.0678 & 71.8675 & 4.6904 & 15.3223 \\
2.1394 & 73.8355 & 4.9648 & 14.8718 \\
2.2135 & 70.5583 & 5.0372 & 14.0076 \\
2.2901 & 66.6173 & 4.8304 & 13.7912 \\
2.3693 & 66.9209 & 3.8831 & 17.2340 \\
    \enddata
    \tablecomments{Throughput-corrected HD 91312 B spectrum extracted from 13 March 2017 data. (a) These spectral channels are likely to be telluric-dominated.
    }
    \label{spectrum_hd91312B}
    \end{deluxetable}

\subsection{Empirical Comparisons to HD 91312 B's Infrared Colors and Spectrum}\label{subsection:empirical}
To empirically constrain HD 91312~B's atmosphere, we first compare its broadband colors to ultra-cool dwarf colors compiled in \citet{Pecaut2013,KenyonHartmann1995} and then its CHARIS spectrum to objects in the Montreal Spectral Library\footnote{\url{https://jgagneastro.com/the-montreal-spectral-library/}} \citep[e.g.][]{Gagne2014}, considering the impact of spatially and spectrally correlated noise \citep{GrecoBrandt2016}.   HD 91312~B's $J - H_{\rm s}$ and $J - K_{\rm s}$ colors converted to the 2MASS photometric system are 0.72 $\pm$ 0.12 and 0.89 $\pm$ 0.12 \citep{Pecaut2013}, respectively, which are matched by M5.5--L1 and K7--M7 field objects at the 1-$\sigma$ level, respectively \citep[see][]{Pecaut2013}.   Its $J$-$L_{\rm p}$ color (0.97 $\pm$ 0.35) does not well constrain its spectral type \citep[later than K4;][]{KenyonHartmann1995}, due to the large NIRC2 photometric errors.

\begin{figure}
    
     \includegraphics[width=0.475\textwidth,trim=10mm 0mm 0mm 0mm,clip]{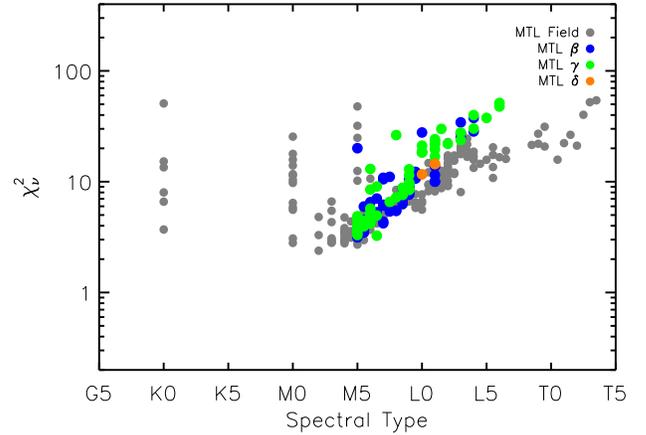}
     \vspace{-0.25in}
    \caption{The $\chi_{\rm \nu}^{2}$ vs. spectral type distribution comparing the spectrum of HD 91312 B to those for objects with spectral types between K0 and T5 in the Montreal Spectral Library.   Field objects are shown as grey dots.   Objects with \textit{very low} ($\delta$), \textit{low} ($\gamma$), and \textit{intermediate} ($\beta$) gravity classifications are shown as  orange, green, and blue dots, respectively. }
    \label{fig:chisq}
\end{figure}

As shown in Figure \ref{fig:chisq}, HD 91312 B's spectrum is best-matched by early-to-mid M dwarfs -- M0 to M6 objects -- which reproduce the H-band peak but otherwise relatively flat spectral shape at $J$ and $K_{\rm s}$.   Later spectral types tend to have K-band shapes that are too bright and peaked to reproduce HD 91312 B's spectrum.   Effective temperatures for M0--M6 dwarfs span $T_{\rm eff}$ $\sim$ 2850--3870 K \citep{Pecaut2013}.   For M dwarfs, the absolute K-band magnitude and bolometric magnitude correlate well.  Adopting the relationship from \citet{Casagrande2008} and assuming a distance of 33.43 $pc$, HD 91312 B's luminosity is $\log_{10}(L/L_{\rm \odot}) = -2.09^{+0.12}_{-0.13}$.

 \subsection{Atmospheric Model Spectra and HD 91312 B}\label{sec:atm-model}
 
For comparison to model stellar spectra, we perform a similar fit of HD 91312 B's observed spectrum to the solar metallicity BT-Settl-CIFIST models \citep{Baraffe2015}. As with the previous BT-Settl models \citep{Allard2012a,Allard2012b}, BT-Settl-CIFIST includes self-consistent cloud formation and sedimentation. However, the latter also includes updated solar abundances and a more detailed treatment of convection \citep{Baraffe2015}. Notably, however, this analysis does not include any constraints on metallicity.

We combine our extracted A-LOCI CHARIS spectrum from 13 March 2017 with the $L_p$ photometric point from NIRC2, ignoring spectral channels that are likely to be telluric-dominated (1.375, 1.422, 1.867, and 1.932 $\mu$m in the CHARIS spectrum). As spectra observed with CHARIS are deconvolved from the Line-Spread Function during extraction \citep{Brandt2017}, the model spectra were convolved with a Gaussian corresponding to the appropriate CHARIS spectral bin width at each wavelength ($FWHM = \lambda / R$ converted to spectral bins) and re-binned for comparison with the observed spectra. As the CHARIS spectral resolution is logarithmic in nature, the FWHM of the convolved Gaussian varied as a function of $\lambda$ across the wavelength range of the CHARIS Broadband filter. To ensure manageable computational time, the same FWHM was used for every 501 model wavelength bins, corresponding to $\sim 0.15\%$ of the width of the CHARIS Broadband filter. Calculation of the expected model flux in the NIRC2 $L_p$ band used a linear interpolation of NIRC2's $L_p$ bandpass curve provided by the SVO Filter Profile Service \citep{Rodrigo2012,Rodrigo2020}.

\begin{figure}
    
     \includegraphics[width=0.475\textwidth,trim=10mm 0mm 0mm 0mm,clip]{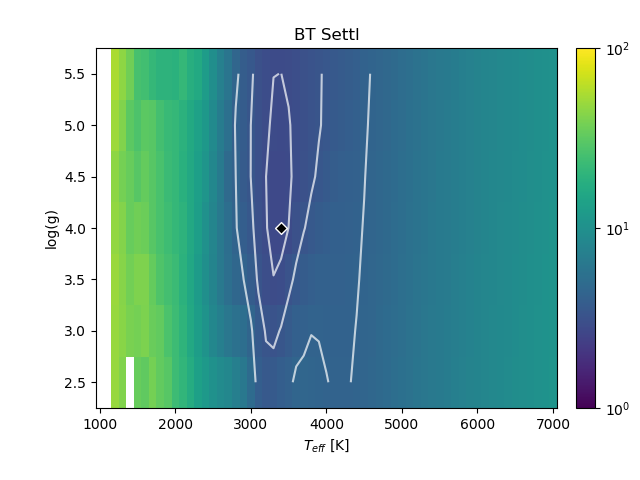}
     \vspace{-0.25in}
    \caption{The $\chi_{\rm \nu}^{2}$ fit of the HD 91312 B spectrum to the BT-Settl-CIFIST models \citep{Baraffe2015} as a function of the model's effective temperature, $T_{eff}$, and $\log(g)$. Fits utilized the Keck $L_p$-band photometry, along with the CHARIS spectrum from the A-LOCI-reduced 13 March 2017 data, excluding four telluric-dominated spectral channels (those at 1.375, 1.422, 1.867, and 1.932 $\mu$m). The best-fit model spectrum (black diamond) has $T_{eff} = 3400$ K and $\log(g) = 4.0$ with a $\chi_{\rm \nu}^2 = 2.75$. Contours show the 1, 3, and 5-$\sigma$ confidence intervals on the $\chi_{\nu}^2$.}
    \label{fig:btsettl-chisq}
\end{figure}

\begin{figure}
    
     \includegraphics[width=0.475\textwidth]{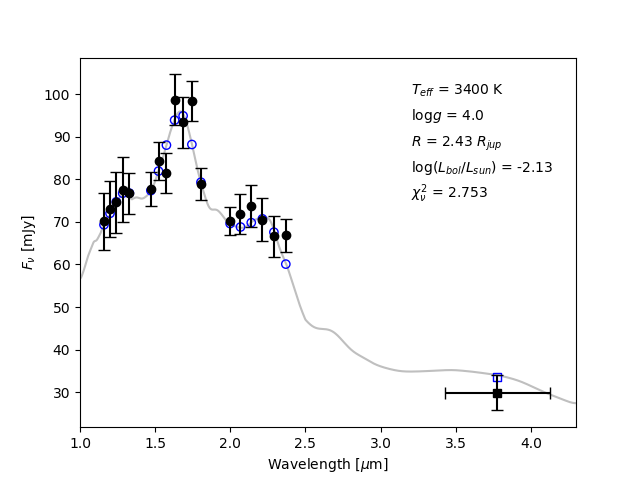}
     \vspace{-0.25in}
    \caption{The HD 91312 B spectrum compared to the best fit BT-Settl-CIFIST model \citep{Baraffe2015}. Model spectrum is shown in gray, smoothed to a spectral resolution of $R=100$ at 2.5 $\mu$m. Binned model spectrum points are denoted as blue open symbols, while the observed HD 91312 B spectrum is shown in black. In both cases, circles indicate points corresponding to the CHARIS spectrum and squares indicate the photometric Lp measurement from Keck. Only spectral points utilized for the model fitting are shown.}
    \label{fig:btsettl-bestspec}
\end{figure}

Once the model spectra have been re-binned to match the observed CHARIS Broadband with NIRC2 $L_p$ photometry, the spectral $\chi^2_{\nu}$ is calculated using the same solution to the correlation model of \citet{GrecoBrandt2016} discussed in Section \ref{subsection:empirical}. The resulting $\chi_\nu^2$ as a function of model $T_{eff}$ and $\log g$ is shown in Figure \ref{fig:btsettl-chisq}. Our spectroscopic/photometric measurements are best fit by the BT-Settl-CIFIST model with $T_{eff} = 3400$ K and $\log g = 4.0$, resulting in $\chi_{\nu}^2 = 2.75$. This best fit model spectrum, shown with the observed data in Figure \ref{fig:btsettl-bestspec}, results in a radius of $R = 2.43 \pm 0.05 R_{Jup}$ and a $\log_{10} (L_{bol}/L_{Sun}) = -2.13 ^{+0.03}_{-0.04}$. Noteably, the $1\sigma$ confidence interval on the $\chi_\nu^2$ fit does not provide strong constraints on $\log g$ (see Figure \ref{fig:btsettl-chisq}), ranging from $\log g = 3.5 - 5.5$. However, the effective temperatures that fall within the $1\sigma$ ($3\sigma$) confidence interval indicate a spectral type of $\sim$ M2 - 4 (M5.5 - K9), with the best-fit $T_{eff} = 3400$ K model corresponding with a spectral type of $\sim$ M3 \citep{Pecaut2013}.

\subsection{Astrometry}\label{sec:astrometry}

\begin{figure*}[ht]
    \centering

\includegraphics[width=0.95\textwidth]{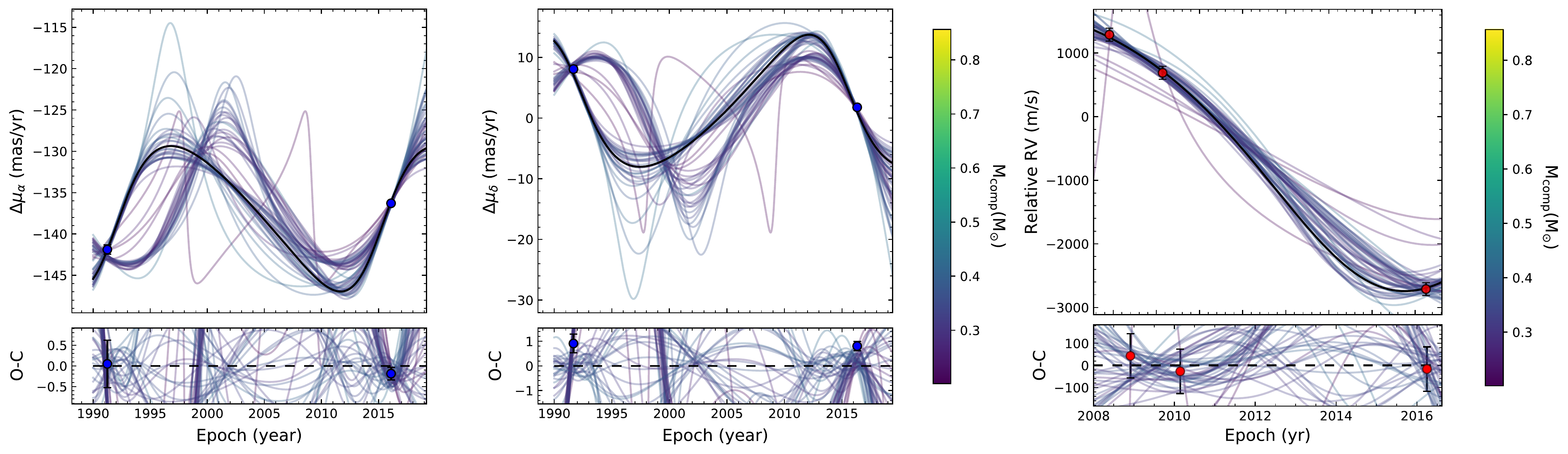}
    \caption{Predicted vs measured proper motion acceleration (left, middle panels) and relative radial velocity (right) for HD 91312 A.  The curves draw from solutions that jointly fit Hipparcos and Gaia astrometry of HD 91312 A, radial-velocity data for HD 91312 A from \citet{Borgniet2019}, and relative astrometry of HD 91312 B from CHARIS and NIRC2.   Blue and red dots denote measurements; the bottom subpanels show residuals from the best fit (black curve) and others color-coded by the dynamical mass of HD 91312 B.}
    \label{fig:dynmass_indirect}
\end{figure*}

\begin{figure*}[ht]
    \centering

\includegraphics[width=0.95\textwidth]{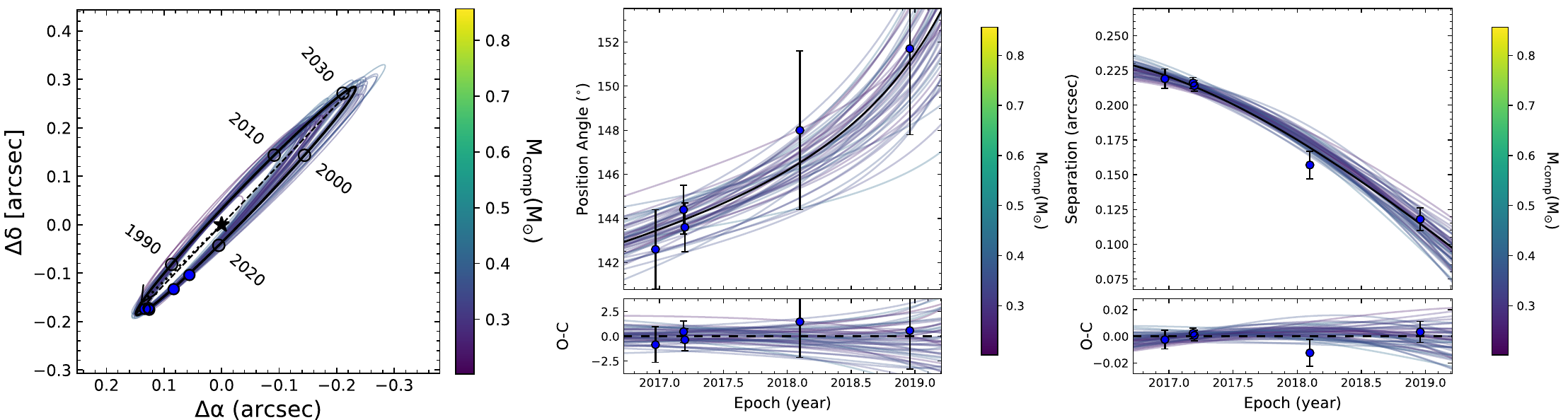}
    \caption{Predicted vs. measured relative astrometry for HD 91312 B.   Curves, data points, and subpanels are the same as in Figure \ref{fig:dynmass_indirect}.}
    \label{fig:dynmass_direct}
\end{figure*}

\begin{figure*}[ht!]
    \centering
    \includegraphics[width=0.95\textwidth]{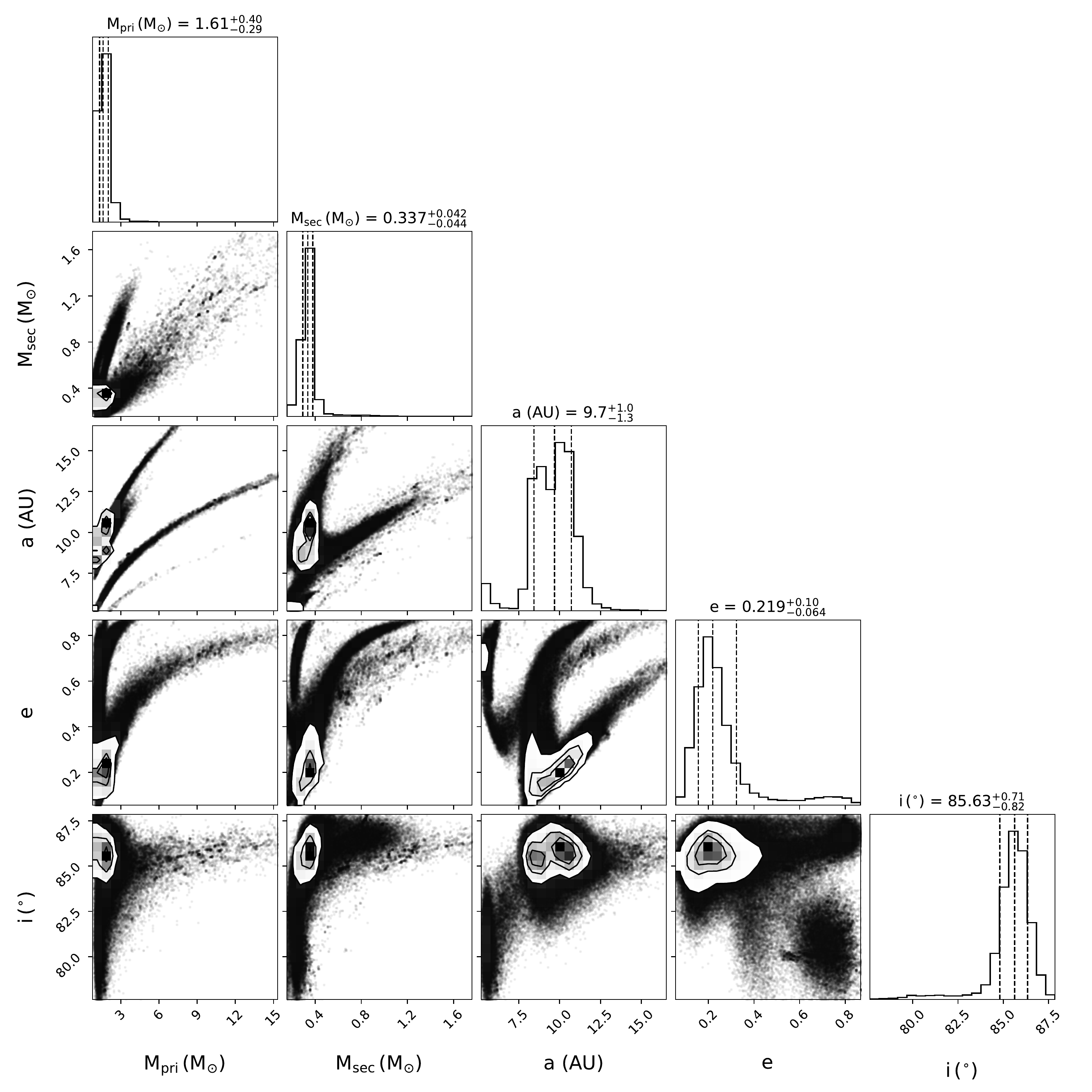}
    \caption{Posterior distributions of selected orbital parameters for HD 91312 B.}
    \label{fig:orbitfit}
\end{figure*}

We use the open-source code \textbf{orvara} \citep{Brandt2021} to fit the low mass companion of HD~91312. The mass and orbit are fit using a combination of the Hipparcos and Gaia Catalogues, radial velocity measurements from \citep{Borgniet2019}, and relative astrometry from SCExAO/HiCIAO, SCExAO/CHARIS and Keck/NIRC2. We fit for the mass of the primary and companion, the semi-major axis, the eccentricity and inclination. We assume that the companion observed is solely responsible for the acceleration observed in between the Hipparcos and Gaia catalogues. For the fit, we assume lognormal priors on mass and semimajor axis, a geometric prior on inclination, a Gaussian prior on parallax using the Gaia EDR3 measurement and its uncertainty, and uniform priors on the remaining parameters. The code analytically marginalizes out parallax, barycenter proper motion, and barycenter radial velocity as nuisance parameters. A full description of \textbf{orvara} and its available parameters can be found it \citet{Brandt2021}. Figure \ref{fig:dynmass_indirect} shows the predicted vs. measured proper motion acceleration and relative radial velocity for HD 91312 A and Figure  \ref{fig:dynmass_direct} shows the predicted vs. measured relative astrometry for HD~91312~B. The companion has a best-fit mass of $0.337^{+0.042}_{-0.044}$~M\textsubscript{\(\odot\)} with a best-fit semimajor axis of $9.7^{+1.0}_{-1.3}$~au and an orbital inclination of $i=85.63^{+0.71}_{-0.82}$ degrees. Figure \ref{fig:orbitfit} shows our posterior distribution of selected orbital parameters.  

\section{Comparison to Stellar Evolution Models}
\label{sec:stellar_evolution_models}

To obtain an age estimate, we compared the derived $T_{eff}$ and luminosity of the HD 91312 AB system to isochrones from several stellar evolution grids. This analysis was performed with the Dartmouth \citep{Dotter2008,Feiden2011}, Yale-Potsdam Stellar Isochrones (YaPSI; \citealt{Spada2017}), MESA Isochrones and Stellar Tracks (MIST; \citealt{Dotter2016,Choi2016}) utilizing the Modules for Experiments in Stellar Astrophysics (MESA) code \citep{Paxton2011,Paxton2013,Paxton2015,Paxton2018}, and the BHAC15 \citep{Baraffe2015} evolutionary models.

\begin{figure*}[ht]
    \centering
\includegraphics[width=0.9\textwidth]{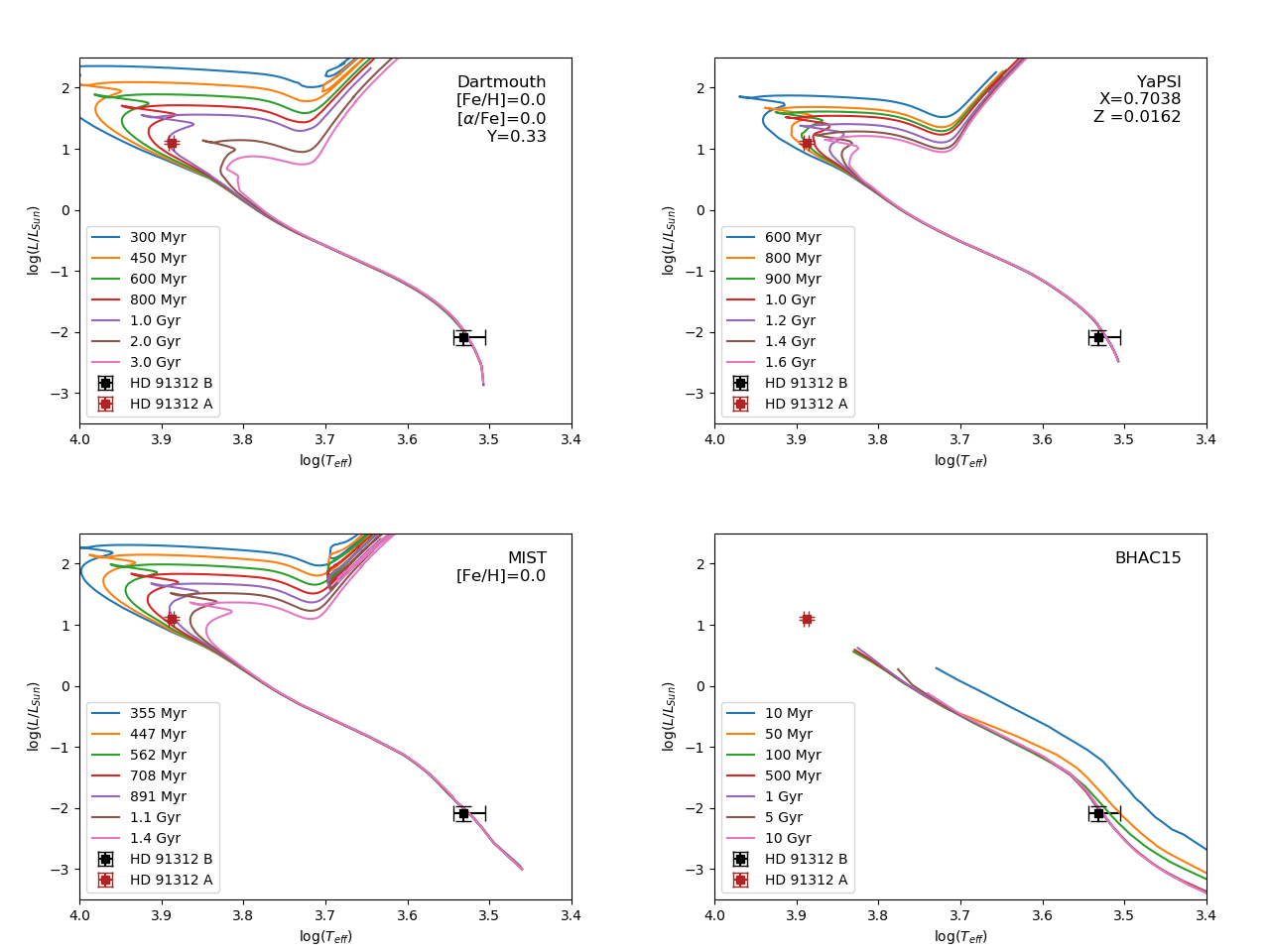}
    \caption{HD 91312 A and B plotted with isochrones at select ages from several stellar evolution grids. The Dartmouth, YaPSI, and MIST isochrones shown are the models closest to solar composition of those that are consistent with the $\log(L/L_{sun})$ and $\log T_{eff}$ of both HD 91312 A and B. The MIST model shown have $v/v_{crit}$ = 0.4. BHAC15 isochrones are not differentiated by composition.}
    \label{fig:isochrones}
\end{figure*}

For the Dartmouth, YaPSI, and MIST models, we consider only isochrones that are simultaneously consistent with the $\log_{10} (L/L_{\rm \odot})$ and $T_{eff}$ of HD 91312 A and B within $1\sigma$. For HD 91312 B, we use the $T_{eff} = 3400_{-200}^{+100}$ K from Section \ref{sec:atm-model} with the empirical $\log_{10} (L/L_{\rm \odot}) = -2.09_{-0.13}^{+0.12}$ (Section \ref{subsection:empirical}). While the empirical and modelled bolometric luminosities derived in Sections \ref{subsection:empirical} and \ref{sec:atm-model} are consistent within their uncertainties, the uncertainties derived from model fitting do not include the uncertainties in the model spectra, themselves. Therefore, we use the empirical bolometric luminosity in this stellar evolution analysis for a more complete representation of the uncertainty. For HD 91312 A, we use $\log T_{eff} = 3.888 \pm 0.003$ and $\log_{10} (L/L_{\rm \odot}) = 1.1040 \pm 0.0178$ from \citet{Zorec2012}. We perform a similar analysis with the BHAC15 models using only the data for HD 91312 B, as HD 91312 A is beyond the mass limits of their models (0.07 - 1.4 $M_{\rm \odot}$; \citealt{Baraffe2015}). Example isochrones from each model at or near stellar metallicity are shown in Figure \ref{fig:isochrones}.

\begin{figure*}[ht]
    \centering
    \includegraphics[width=0.95\textwidth]{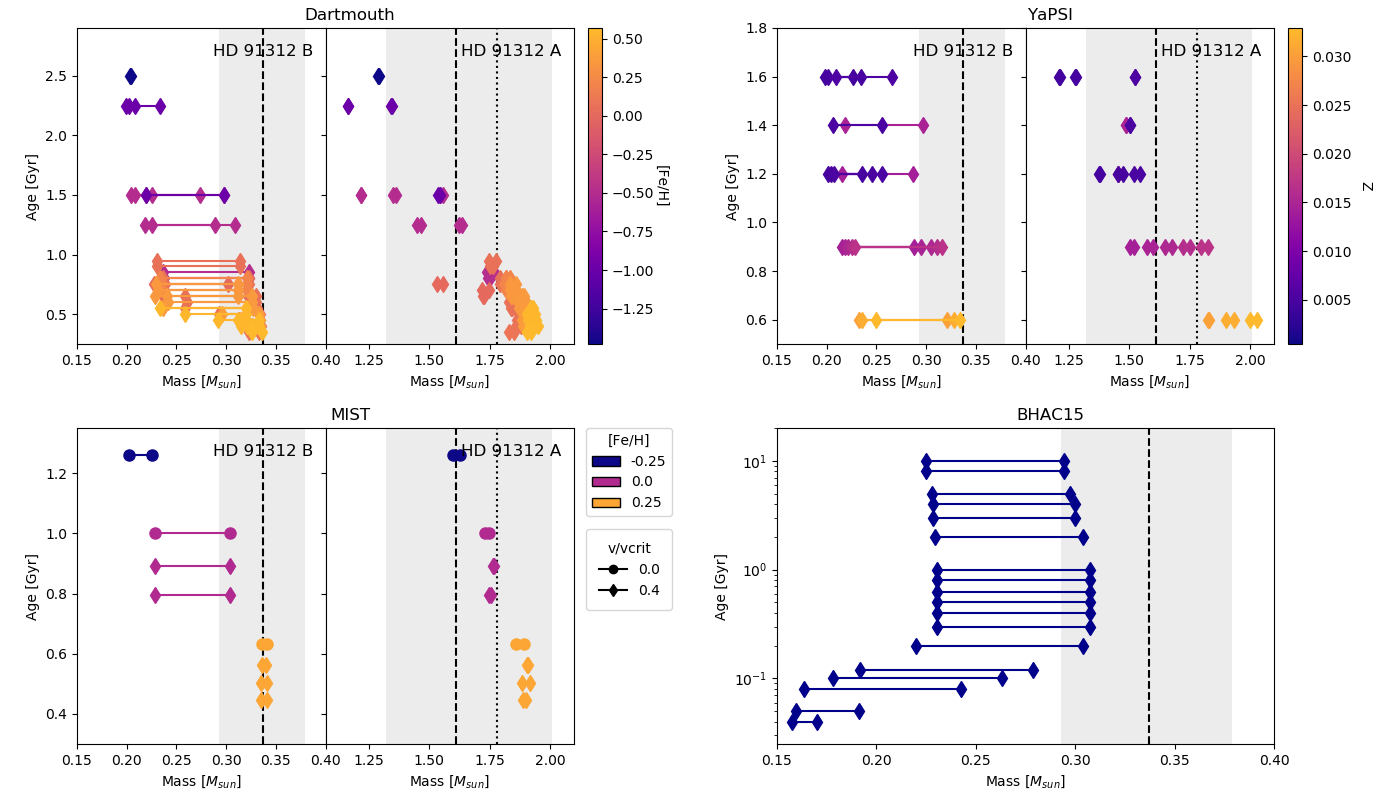}
    \caption{System age vs. mass as derived from several stellar evolution models. For the Dartmouth, YaPSI, and MIST models, masses shown were derived from luminosity-$T_{eff}$ isochrones consistent with both HD 91312 A and B within the $1-\sigma$ uncertainties. Mass ranges are shown for both sources. The \citeauthor{Baraffe2015} models do not extend to the mass of HD 91312 A, so the ages shown are derived from isochrones consistent with HD 91312 B, alone. The dashed black line and gray shading indicate the astrometric masses derived in Section \ref{sec:astrometry} and their uncertainty, respectively. The dotted black line indicates the HD 91312 A mass from \citet{Zorec2012}.}
    \label{fig:agemass}
\end{figure*}

For each isochrone that is simultaneously consistent with HD 91312 A and B (or B alone, in the case of the BHAC15 models), we estimate the range of masses for each source that lie along the isochrone within $1\sigma$ of the measured $\log_{10} (L/L_{\rm \odot})$ and $T_{eff}$ for the source. Figure \ref{fig:agemass} provides a more detailed view of the allowed mass range for each source as a function of isochrone age and, in the cases of the Dartmouth, YaPSI, and MIST models, metallicity. 

\begin{figure*}[ht]
    \centering
    \includegraphics[width=0.75\textwidth]{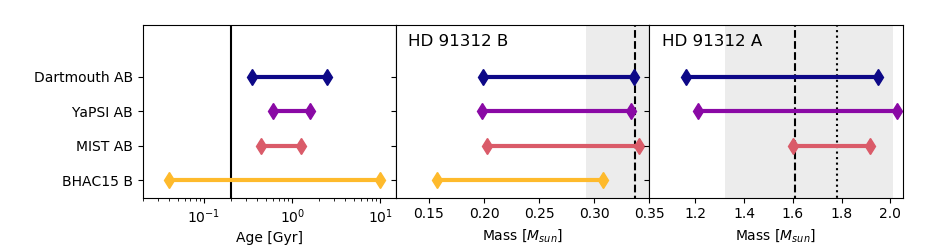}
    \includegraphics[width=0.75\textwidth]{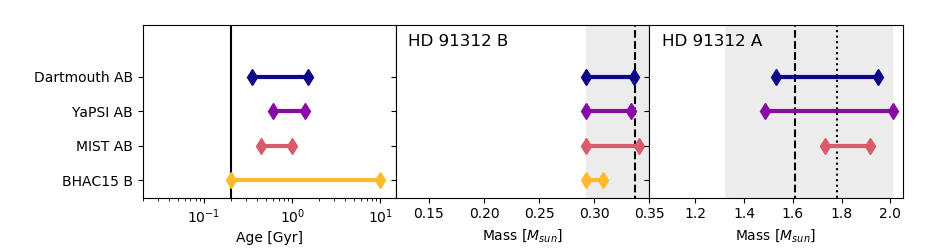}
    \caption{Overview of the consistent age and mass ranges from all isochrone fits with no mass constraints (top) and after constraining by the astrometric masses in Section \ref{sec:astrometry} (bottom). Both figures give the range of allowed stellar ages (left) as well as the mass ranges for HD 91312 B (center) and A (right) from all isochrone fits. Model name is labeled to the left, along with whether the fits were performed with both 91312 A and B or B alone. The solid black line denotes the age of HD 91312 A as derived by \citet{Rhee2007}. The dashed black lines with shading in the center and right figures indicate the astrometric mass and uncertainty, respectively derived in Section \ref{sec:astrometry}. The dotted black line shows the mass of 91312 A from \citet{Zorec2012}.}
    \label{fig:evol_model_ranges}
\end{figure*}

When combined with the astrometric mass for HD 91312 B derived in Section \ref{sec:astrometry}, the system becomes inconsistent with the lowest metallicity Dartmouth ([Fe/H]$ < -1.0$), YaPSI (Z$ < 0.01$) models, and MIST ([Fe/H]$ < 0.0$) models. Figure \ref{fig:evol_model_ranges} shows the full range of ages and masses for each source allowed by each of the four stellar evolution grids, both before and after constraining the mass ranges for HD 91312 A and B to be within $1\sigma$ of the astrometric masses. With no mass constraints, the $1\sigma$ age ranges given by each model are $0.35 - 2.5$ Gyr (Dartmouth AB), $0.6 - 1.6$ Gyr (YaPSI AB), $0.45 - 1.3$ Gyr (MIST AB), and $0.04 - 10$ Gyr (BHAC15 B). Imposing the mass constraints reduces the allowed age ranges to $0.35 - 1.5$ Gyr (Dartmouth AB), $0.6 - 1.4$ Gyr (YaPSI AB), $0.45 - 1.0$ Gyr (MIST AB), and $0.2 - 10$ Gyr (BHAC15 B).

\section{Conclusions \& Discussion}
In this work, we directly image and characterize a low-mass stellar companion to the A7 star HD 91312 with SCExAO/CHARIS, SCExAO/HiCIAO, and Keck/NIRC2.   The presence of HD 91312 B is consistent with a long-term radial velocity trend seen by ground-based Doppler surveys \citep{Borgniet2019} and space-based precision astrometry missions \citep{GAIA2018,Brandt2018}.

HD~91312~B's spectrum is best matched by early to mid M dwarf spectra.   Modeling the SCExAO/CHARIS spectrum and $L_{\rm p}$ photometry from Keck/NIRC2 using the BT-Settl-CIFIST models yields a best fit to a 3400$_{-200}^{+100}$ K object with a log(g)=4.0$_{-0.5}^{+1.5}$, consistent with empirical comparisons.
. 
Combining relative astrometry of HD 91312 B with radial-velocity measurements and absolute astrometry of the star yields a dynamical mass for the companion of $0.337^{+0.042}_{-0.044}$~M\textsubscript{\(\odot\)}, also consistent with the results from spectral fitting. To obtain an age estimate of the HD~91312~AB system, we derived $T_{eff}$ and luminosity and compared these to isochrones from several stellar evolution grids. While specific constraints vary, the Dartmouth, YaPSI, and MIST models for the 91312 AB system predict an age between $\sim 0.35 - 1.5$ Gyr when including astrometric mass, and are, in general inconsistent with the lowest metallicity models.

The combination of direct imaging with indirect methods provides a clearer picture of the HD 91312 system.  Absent dynamical mass information from astrometry and radial-velocity, the mass for HD 91312 B would have to have been inferred from luminosity evolution models assuming a system age, which is not tightly constrained.   The inclusion of data from these indirect methods, however, yields a mass precision of $\sim$10\%.   Finally, this result with other companions discovered from our Hipparcos-\textit{Gaia} direct imaging survey reinforce the value of using astrometry to screen for promising direct imaging targets \citep{Currie2020b, Steiger2021}.   The revised version of HGCA uses \textit{Gaia}-eDR3 astrometry, which is about a factor of 3 more precision than that in the DR2 release (Brandt et al. 2021, submitted), making the catalogue more sensitive to the presence of jovian companions.   The first joint direct imaging + astrometry discovery of an exoplanet is likely only a matter of time.

\acknowledgments
\indent We thank the anonymous referee for helpful comments that improved the quality of this paper.
\indent The authors wish to acknowledge the very significant cultural role and reverence that the summit of Maunakea holds within the indigenous Hawaiian community.  We are most fortunate to have the opportunity to conduct observations from this mountain.

\indent We wish to acknowledge the critical importance of the current and recent Subaru and Keck Observatory daycrew, technicians, telescope operators, computer support, and office staff employees.  Their expertise, ingenuity, and dedication is indispensable to the continued successful operation of these observatories. 
\\
\indent T.C. was supported by a NASA Senior Postdoctoral Fellowship and NASA/Keck grant LK-2663-948181.  We thank the Subaru and NASA Keck Time Allocation Committees for their generous support of this program. \\
\indent The development of SCExAO was supported by JSPS (Grant-in-Aid for Research \#23340051, \#26220704 \& \#23103002), Astrobiology Center of NINS, Japan, the Mt Cuba Foundation, and the director's contingency fund at Subaru Telescope.  CHARIS was developed under the support by the Grant-in-Aid for Scientific Research on Innovative Areas \#2302.   Some of the data presented herein were obtained at the W. M. Keck Observatory, which is operated as a scientific partnership among the California Institute of Technology, the University of California and the National Aeronautics and Space Administration. The Observatory was made possible by the generous financial support of the W. M. Keck Foundation.\\
\indent K.W. acknowledges support from NASA through the NASA Hubble Fellowship grant HST-
HF2-51472.001-A awarded by the Space Telescope Science Institute, which is operated by
the Association of Universities for Research in Astronomy, Incorporated, under NASA
contract NAS5-26555.\\
\indent M.T. is supported by JSPS KAKENHI grant Nos.18H05442, 15H02063, and 22000005. \\
\indent This work was supported by JSPS Grants-in-Aid for Scientic Research, 17K05399 (E.A.).

\newpage
\bibliographystyle{aasjournal}                                                             
\bibliography{Arxiv_hd91312}

\end{document}